\documentclass[preprint,12pt,authoryear]{elsarticle}

\usepackage{amsmath}
\usepackage{amssymb}
\usepackage{url}

\journal{Advances in Space Research}

\begin{document}
\begin{frontmatter}

\title{Effects of the Great American Solar Eclipse on the lower ionosphere observed with VLF waves}

\author[1]{Rok Vogrin\v{c}i\v{c}}
\ead{rok.vogrincic@fmf.uni-lj.si}

\author[2,3]{Alejandro Lara\corref{cor1}}
\ead{alara@igeofisica.unam.mx}
      
\author[4]{Andrea Borgazzi}
\ead{aborgazzi@herrera.unt.edu.ar}
 
 \author[5]{ Jean Pierre Raulin}
\ead{raulin@craam.mackenzie.br}

\address[1]{Faculty of Mathematics and Physics, University of Ljubljana, Jadranska ulica 19, 1000 Ljubljana, Slovenia}
\address[2]{Instituto de Geof\'isica, Universidad Nacional Aut\'onoma de %
  M\'exico (UNAM), CdMx,  04510, M\'exico }
\address[3]{The Catholic University of America, Washington D.C, 20064, USA.}
\address[4]{Facultad de Ciencias Exactas y Tecnol\'ogicas (FACET), Universidad Nacional de Tucum\'an (UNT), Tucum\'an, 4000, Argentina}
\address[5]{CRAAM - Centro de Radioastronom\'ia e Astrof\'isica Mackenzie Escola de Engenharia, Universidade Presbiteriana Mackenzie Rua da Consolac\~ao, 896,  CEP 01302-907 S\~ ao Paulo, Brazil}

\cortext[cor1]{Corresponding author}

\begin{abstract}

The altitude of the ionospheric lower layer (D-region) is highly
influenced by the solar UV flux affecting in turn, the propagation of
Very Low Frequency (VLF) signals inside the waveguide formed between
this layer and the Earth surface. The day/night modulation observed in
these signals is generally used to model the influence of the solar 
irradiance onto the D-region. Although, these changes are relatively slow and the transitions are ``contaminated''  by  mode coalescences. In this way, a rapid change of the solar irradiance, as during a solar eclipse, can help to understand the details of the energy transfer of the solar radiation onto the ionospheric D-layer. 
Using the "Latin American VLF Network" (LAVNet-Mex) receiver station in Mexico City, Mexico,
we detected the phase and amplitude changes of the VLF signals transmitted by the NDK station at  25.2 kHz  in North Dakota, USA during the August 21, 2017  solar eclipse.
As the Sun light was eclipsed, the rate of ionization in the
ionosphere (D-region) was  reduced and the effective reflection height
increased, causing a considerable drop of the phase and amplitude of
the observed VLF waves. The corresponding waveguide path is  3007.15
km long and crossed almost perpendicularly the total
eclipse path.
Circumstantially, at the time of the total eclipse a C3 flare took place allowing us to isolate the flare flux from the background flux of a large portion of the disk.
In this work we report the observations and  present a  new  model of
the ionospheric  effects of the  eclipse
and  flare. The model is based on a detailed 
setup of the degree of Moon shadow that affects the entire
Great
Circle Path (GCP).
This relatively simple model, represents a new
approach to obtain a good measure of the reflection height variation
during the entire eclipse time interval.
During the eclipse, the maximum phase variation was -63.36$^{\circ}$ at 18:05 UT which,
according to our model, accounts for a maximum increase of the reflection
height of  9.3 km.

\end{abstract}


\end{frontmatter}

\section{Introduction}

The propagation of very low frequency (VLF) radio signals  within the
Earth-Ionosphere waveguide is strongly affected by changes of the solar
UV, e. g., during the day/night slow variation or  the rapid changes
during a solar flare.
The upper boundary of this waveguide is the D-region of the ionosphere,
starts at $\sim$ 70-80 km above the surface of the Earth, and is
formed by the  Lyman-$\alpha$ radiation from the Sun which ionizes
 molecular Nitrogen and Oxygen;  Nitric oxide; and  various atoms such
 as Sodium and Calcium \citep{Nicolet60}.

During a solar eclipse, the  UV solar 
flux decreases and  consequently, the rate of ionization in the D-region is
strongly reduced, causing an elevation of the
effective height of the reflecting ionospheric layer.
Thus the conditions in the lower ionosphere approach to those observed during the night-time~\citep{MendesDaCosta:1995, Tereshchenko:2015}.
Making solar eclipses very helpful natural setups to study the ionization dynamics  of the lower ionosphere.

The effects of solar eclipses on the amplitude and phase of  VLF signals and the consequent estimation of the ionospheric reflection height at the times of maximum phase of the eclipse, have been investigated  by several authors,
using different techniques \citep[e. g. ][]{MendesDaCosta:1995, Clilverd:2001, Guha:2012, Kaufmann:1968, De:2010, De:1997}, as instance:
using the Long Wave Propagation Capability waveguide code (LWPC) to calculate the change of ionospheric reflection height,
from its
unperturbed value  of  $H'=71$ km 
\citet{Clilverd:2001} found 
that  the 
maximum effects during the  August 11, 1999 eclipse  occurred when 
the effective height parameter $H'$ was 79 km, on a GCP of 1245 km at a steep angle with respect to the totality path of the eclipse.
Also, these effects  were investigated 
through the measurements of VLF sferics by \citet{Guha:2012} who
calculated  increases of the reflection height of $\sim 4.85$ km on a
GCP $>$ 10,000 km for the  July 22, 2009 eclipse; 
and
$\sim 5.14$ km on a GCP $<$ 10,000 km for the January 15, 2010 eclipse. 
By comparing the eclipse with the day/night phase changes, 
\cite{MendesDaCosta:1995} estimated the maximum phase retardation
as 43$\%$ of the total diurnal average phase change, which 
corresponds to a  rise of the VLF effective reflection height of  6.18 km on a GCP of 2820 km. 
Furthermore, \cite{De:2010} 
via a quantitative relationship between the phase delay and the reflection height change, a  rise of the VLF reflection height of around 3.75 km on a GCP of 5761 km
during the August 1, 2008 eclipse.
%
Recently, 
\cite{2019JGRA..124..616V} estimated the
reflection height increase of  3 km on a GCP of 4800 km for the
July 22, 2009 eclipse; and also \cite{2012P&SS...73..310P}  reported a
3 km increase on a GCP of 5700 km for the January 15, 2010 eclipse. 

Besides ground observations, rocket measurements of the total electron
density have been performed during solar eclipses,
reporting effective VLF reflection height increasings of $\sim 9$ and
$\sim  8$ km
during the eclipses observed on 
November 12, 1966 in  Brazil and March 7, 1970 in Virginia, respectively \citep{Clilverd:2001}.

It is worth noting that during a total solar eclipse the magnitude of
VLF signal amplitude decrease depends on GCP length and its
orientation with totality path, time of eclipse, and the eclipse
magnitude \citep{2019JGRA..124..616V}. Therefore one can expect slightly different decreases in signal amplitude for propagation paths of similar lengths. 

In this paper we present the VLF observation of the ``Great American''
eclipse of August 21, 2017 done by the LAVNet-Mex receiver station
 \citep{Borgazzi:2014} in Mexico City, Mexico (Sec. \ref{sec:exp}).
We studied the eclipse effects on the phase and amplitude of the NDK transmitter
signal (25.2 kHz, from La Moure, ND, USA) over the correspondent  path length of
3007.15 km (Sec. \ref{sec:obs}). In particular we  present a phase
deviation model that includes eclipse effects, such as shortening of
the propagation path between transmitter and receiver, and increases
in the effective reflection height of the ionosphere (Sec. \ref{sec:model}).     
We also have modeled the effects of a C3 flare occurred at the time of
the maximum occultation (Sec. \ref{sec:flare}). Finally our summary is
in Sec. \ref{sec:summary}.

\section{Experimental setup} \label{sec:exp}

For this study, 
we used the measurements of phase and amplitude of the VLF waves detected with the Latin American Very Low Frequency Network at Mexico (LAVNet-Mex) receiver station, which
operates at the frequency range of 10-48 kHz and is located at the
Geophysics Institute of the National Autonomous University of Mexico
(UNAM), at 99$^{\circ}$ 11$'$ W, 19$^{\circ}$ 20$'$ N,
\citep{Borgazzi:2014}. LAVNet-Mex is formed by two loop-type antennas;
each one has a very low noise preamplifier in differential input
configuration, and uses a commercial high quality sound card as a
digitizer. The system bandwidth is 40 kHz centered at 30 kHz, the
voltage gain is 51.88 dB and the common-mode rejection ratio is 74.83
dB. The two wire-loop antennas (N-S and E-W configuration) have 100
turns, mounted on aluminum square frame with 1.8 m per side, which
gives around 324 m$^{2}$ of effective area, \citep{Borgazzi:2014}. We
achieved successful phase measurements with the use of a compact GPS
receiver through a one pulse per second (PSS) signal. Then, the Sound card's crystal-clock signal is locked to the GPS internal clock (PPS). The resulting signal phase has a precision of less than $\sim 1^{\circ}$ \citep{Raulin:2010}.  

\section{The Great American Solar Eclipse} \label{sec:eclipse}
\begin{figure}
\begin{center}
\includegraphics[scale=0.52]{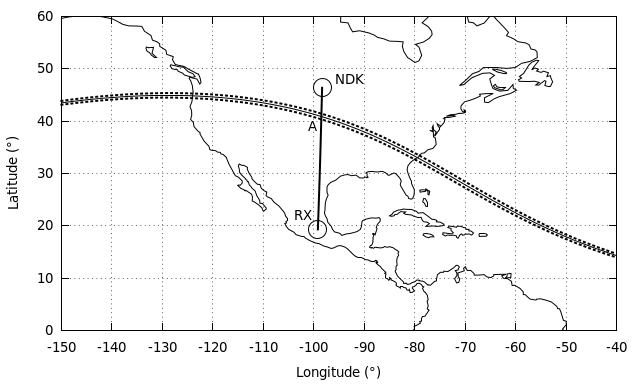}
\caption[f1]{Map of the path of the August 21, 2017 eclipse. The
  continuous line represent the center of the totality, whereas the  lower
  and upper limits are marked with  dashed lines. The great circle
  from the transmitter (NDK) to the receiver (RX) in Mexico City,
  Mexico is marked with the thick full line. The point A on the map
  represents the location of the maximum of the eclipse on our propagation path, the maximum phase of the eclipse at point A occurred at 18:00 UT.}
\label{fig:Figure 1}
\end{center}
\end{figure}

The Great American solar eclipse (August 21, 2017)  began  at the
North Pacific Ocean; the inland totality first occurred at $\sim$  17:17
UT in Oregon; the last contact with the mainland occurred at around
18:48 UT in South Carolina; and finally, the eclipse ended at the North Atlantic Ocean. 
The maximum duration of totality was 2m40s. This eclipse provided us with a rare opportunity to study the effects on the propagation of VLF waves within the Earth-ionosphere waveguide because the path of totality crossed the propagation path between the transmitter and the receiver. 
The path and locations of the NDK transmitter and the LAVNet-Mex receiver (RX) with respect to the path of totality on August 21, 2017, are shown in Figure \ref{fig:Figure 1}. The great circle path between the transmitter and the receiver is 3007.15 km long. The continuous line represents the center of the shadow, while the dashed lines represent the lower and the upper limits of the totality shadow. Point A (40$^{\circ}$ 54$'$ N, 98$^{\circ}$ 27$'$ W) represents the location of the  maximum eclipse on our propagation path, 
at 18:00 UT. 
We note that NDK-RX path crosses the path of totality almost perpendicularly. At the receiver site 
the maximum obscuration was $\sim 27\%$ at around 18:20 UT, while at the transmitter site NDK (46$^{\circ}$ 22$'$ N, 98$^{\circ}$ 20$'$ W) the obscuration was $\sim 83\%$  at around 17:57 UT.

\section{VLF Observations}\label{sec:obs}
\begin{figure}
\begin{center}
\includegraphics[scale=0.25]{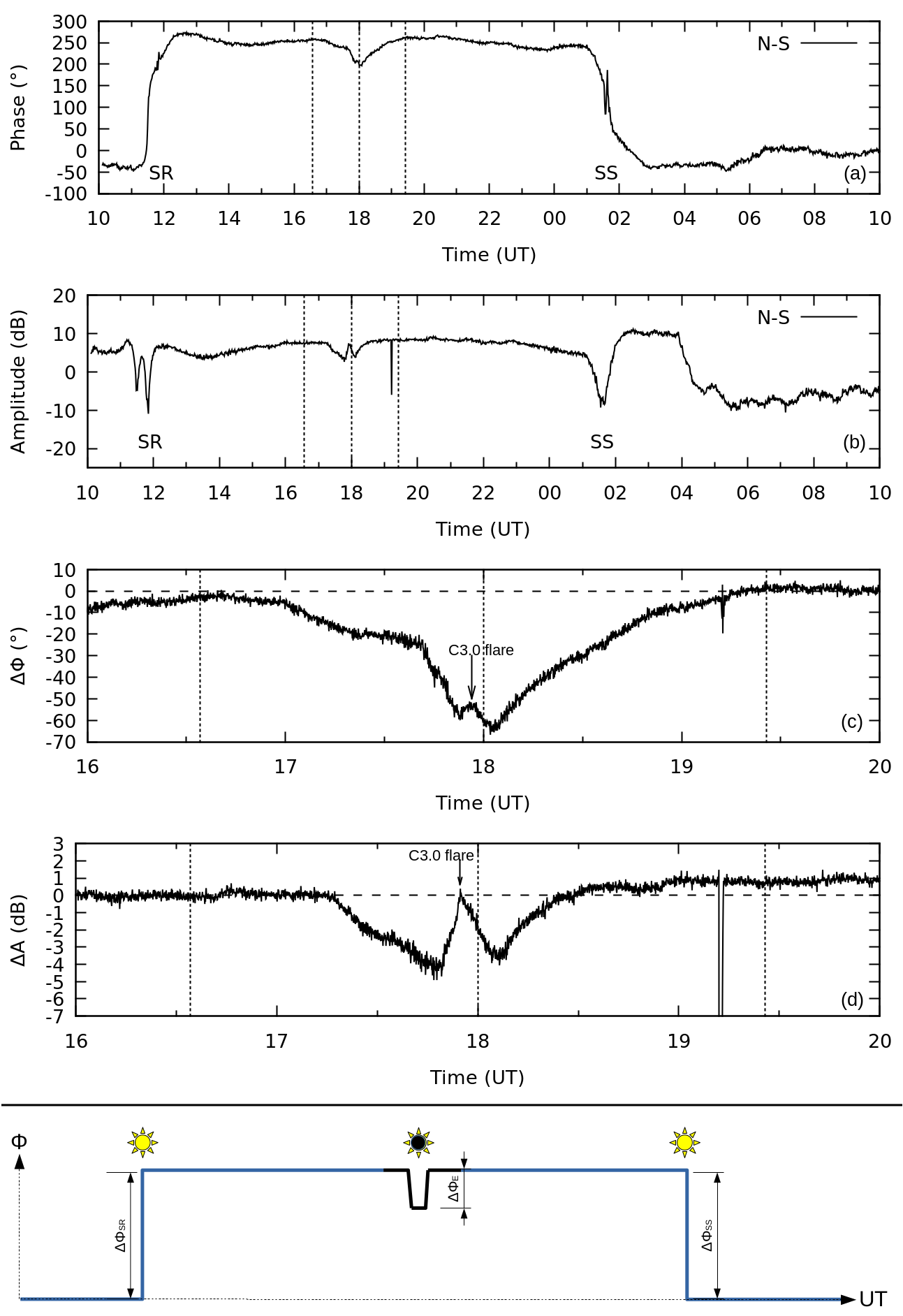}
\caption[f2]{
Phase (a)  and  amplitude (b) variations of the VLF signal along
NDK-RX path, from 10:00 UT on August 21, 2017 to 10:00 UT next day,
of the  N-S  antenna. We marked the sunrise (SR) and the sunset (SS)
where the phase changes abruptly, according to the diurnal pattern. 
A detailed view of the phase (c) and the  amplitude (d)  at the time of the eclipse (16:00 to 20:00 UT). The vertical dashed lines indicate  the beginning (16:34 UT), the maximum (18:00 UT) and the end (19:26 UT) of the eclipse along our propagation path.  The maximum of the C3.0 solar flare, at 17:57 UT is marked.
The bottom panel schematically represents the diurnal phase change. $\Delta \Phi_E$ represents a phase change due to the eclipse, while $\Delta \Phi_{SR}$ and $\Delta \Phi_{SS}$ represent a phase change between the night and the day, one at sunrise time  (SR) and one at sunset time (SS). }
\label{fig:Figure 2}
\end{center}
\end{figure}

In Figure \ref{fig:Figure 2} we show the phase (a) and the amplitude
(b) variation of the VLF signals propagating along NDK-RX path, from
10:00 UT on August 21, 2017 to 10:00 UT next day, for the N-S antenna. We marked sunrise (SR) and sunset (SS) according to the diurnal pattern where the phase increases/decreases abruptly, 
at $\sim$ 12:00 UT and  $\sim$ 01:30 UT the next day,
respectively. The night-time value of phase is $\sim 0^{\circ}$, while
the unperturbed daytime value is $\sim 260^{\circ}$.
The amplitude shows
similar patterns of sunrise/sunset with an  unperturbed daytime value of
$\sim 8$ dB.
The vertical dashed lines indicate the beginning
(16:34 UT), the maximum (18:00 UT) and the end (19:26 UT) of the
eclipse along our propagation path.

The bottom panel in Figure \ref{fig:Figure 2} represents a rough
schematic of the diurnal phase change. We are interested in the ratio
of the phase change due to the solar eclipse ($\Delta \Phi_E$) and the
phase change between the night and day during the sunrise ($\Delta
\Phi_{SR}$) and sunset ($\Delta \Phi_{SS}$). The maximum phase change
during the eclipse accounted for $\sim$25$\%$ of the total diurnal
average phase change. Assuming a total diurnal change (at the
reference height) of  $\sim$ 20-30 km and an upper limit of the
daytime D-region of 90 km of altitude, the eclipse  phase change represents a rise
in the effective reflection height of the ionosphere of $\sim$ 5-8 km.
In Table \ref{tab:Table 1} we present the measured phase change, the ratios of eclipse with respect to day-night phase change and the estimated change in the effective reflection height of the ionosphere.

In panels (c) and (d) of Figure \ref{fig:Figure 2}, we present a
detailed view of the phase and the amplitude changes during the time
of the solar eclipse (16:00 - 20:00 UT).  
For the sake of simplicity, 
we shifted to zero  the values of the phase and the amplitude during unperturbed times. 
The phase  reaches a minimum of -63.36$^{\circ}$ at 18:05 UT. 
We note that 
at the time of the eclipse
a C3.0 solar flare  occurred, with a maximum at 17:57 UT (in X-ray flux). The flare was powerful enough to shift the phase towards positive values thus distorting the eclipse pattern. 
In this way,
the real minimum should be occurred a few minutes prior to the measured one, at 18:00 UT, which is the time of totality on our propagation path (point A in Figure \ref{fig:Figure 1}). 
The change in amplitude reaches a minimum of -4.90 dB at 17:47 UT,
however the effect of the C3.0 flare on amplitude is more profound
than the case of the phase.

Differences between amplitude and phase time-behavior are regularly
seen at sunrise and sunset \citep{wait1968mode, 1983ZaMM...63..281L,davies65}  as clearly seen at 11:30 UT (sunrise) and at 03:30 UT (sunset) in panels (a) and (b) of Figure 2. These are due to the different  response of the phase delay and amplitude to the modal conversions due to the sudden changes of the height and surface characteristics of the waveguide \citep{wait1968mode}.
Similar differences are seen during the eclipse (panels (c) and (d) of
Figure 2). The phase delay is more sensitive than the amplitude to the
small changes of the Sun illumination before the second contact.
As the quiescent D-layer uplifts at the beginning of the eclipse, the phase velocity of the first-order mode decreases and so does the phase shift. The change of the phase-shift slope may occur when the wave-guide reaches an altitude high enough to allow a significant second-order mode propagating along with the first-order mode causing a further reduction of the phase velocity.

\begin{table}
\begin{center}
\begin{tabular}{c|c|c|c }
\hline
 & \textbf{$\Delta \Phi$($^{\circ}$)} & \textbf{Ratio}  & \textbf{$\Delta$z(km)}\\ 
\textbf{Sunrise (SR)} & 300.15  $\pm$  4.21  & 0.24  $\pm$  0.01 & (4.8-7.2) $\pm$ 0.1\\
\textbf{Sunset (SS)} & 273.40  $\pm$  3.21 & 0.27  $\pm$  0.01 & (5.4-8.1) $\pm$ 0.1\\ 
\textbf{Eclipse (E)} & 72.84  $\pm$  2.60 & & \\ 
\hline
\end{tabular}
\end{center}
\caption{The measured phase  (column 2, $\Delta \Phi$) during the sunrise (SR), and sunset (SS); and the corrected value between noon and the eclipse minimum (E). 
The ratio between phase changes and the expected  change of the ionospheric reflection layer $\Delta$z, are in columns 3 and 4.}
\label{tab:Table 1}
\end{table}

\section{The Eclipse Model} \label{sec:model}

In this section we present a model of the phase deviation during the
eclipse. We consider  effects such as the equivalent shortening of the
propagation path between transmitter and receiver, and increasing of
the effective ionospheric reflection height. We start  with the
equation formulated by \citet{1959ITAP....7..154W},
see also 
\citet{1972JATP...34..255D} and \citet{ Muraoka:1977}:
\begin{equation}
\label{eq:Equation 1}
\Delta \Phi = 360^{\circ}\frac{d}{\lambda}\left(\frac{1}{2a}+\frac{\lambda^2}{16z^3}\right)\Delta z,
\end{equation}
where $\Delta \Phi$ is the phase delay, $d$ is the distance between
transmitter and receiver, $\lambda$ is the wavelength of the VLF wave,
$a$ is the radius of the Earth, $\Delta z$ is the variation of the reflection
height and $z$ is the daytime altitude of the ionosphere. In this
analysis we used z =
70.5 km as  proposed by \cite{Thomson:2010}. We note that Equation
\ref{eq:Equation 1} is valid for a propagation path of distance $d \ge
3000~  km$
illuminated by the Sun and 
our propagation  path is in the lower limit, 
even during the
eclipse when only a small part of the path is totally obscured by the
Moon.

\begin{figure}
\begin{center}
\includegraphics[scale=0.50]{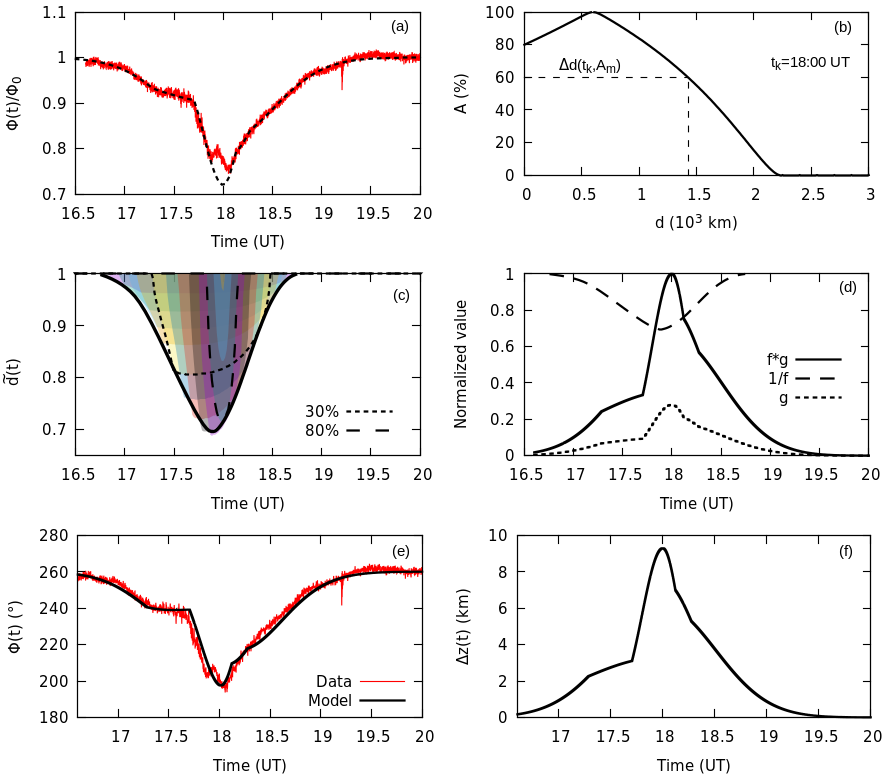}
\caption[f4]{(a) The normalized measured  phase $\Phi(t)/\Phi_0$ (red)
  was approximated by the addition of  6-Gaussian profiles (dashed
  black line). 
 (b) Percentage of 
the covered solar disk as a function of the distance along 
the NDK-RX  propagation path at  18:00 UT (continuous line) the
horizontal dashed line represents the distance where the solar disk
was 60\% covered at this specific time.
(c) The iso-curves of the amount of \% covered as a function of time
(discontinuous lines to mark 30\ and 80\% and colored contours mark
other percentage levels) and the resulting envelope  ($\widetilde{d(t)}$) for the entire time range  
 (thick solid line). 
 (d) The discrete convolution 
between  $f = 1/\widetilde{d(t)}$ (dashed line) and $g = \widetilde{\Phi(t)}$, (dot-dashed line). 
The convolution (continuous line) is proportional to $C_2 \Delta z(t)$.
(e) Our model (black) which best fits the measured phase (red) when $b = 9.3 \pm 0.1$ km.
(f) Change of the ionospheric reflection height during the eclipse. }
\label{fig:model}
\end{center}
\end{figure}

The 
steps of the eclipse modelling are illustrated in Figure
\ref{fig:model}. First,
in order  to filter out the effect of the C3.0 solar flare,
we fitted the observed phase profile, at the time of totality  (without considering  the time of the flare) with
a model constructed by the 
addition of Gaussian curves and 
due to the asymmetric nature of the phase time profile, the best fit
was achieved by adding six Gaussian models (dashed black line in panel a).

Next, we propose that the illuminated propagation path distance is a
function of time, since the fraction of the path covered by the
eclipse shadow changes rapidly. Thus we can write Equation \ref{eq:Equation 1} as: 
\begin{equation}
\label{eq:Equation 2}
\Delta \Phi(t) = C_1 \Delta d(t) \Delta z(t),
\end{equation}
where $C_1 =
360^{\circ}\frac{1}{\lambda}(\frac{1}{2a}+\frac{\lambda^2}{16z^3})$ is
a constant, $\Delta  d(t)$ is a time function of the propagation path
distance and $\Delta z(t)$ is a time function of the reflection height
variation. We can rewrite Equation \ref{eq:Equation 2}  in terms of
relative changes as:
\begin{equation}
\label{eq:Equation 3}
\widetilde{\Phi(t)} = C_2 \widetilde{d(t)}\Delta z(t),
\end{equation}
where the relative phase change is: 
\begin{equation}
\label{eq:Equation 4}
\widetilde{\Phi(t)} = \frac{\Delta \Phi(t)}{\Phi_0} = \frac{\Phi_0-\Phi(t)}{\Phi_0} = 1 -\frac{\Phi(t)}{\Phi_0};
\end{equation}
and the relative propagation path distance is:
\begin{equation}
\label{eq:Equation 5}
\widetilde{d(t)} = \frac{d(t)}{d}= \frac{d- \Delta d(t)}{d}=1 - \frac{\Delta d(t)}{d};
\end{equation}
$C_2 = C_1 d/\Phi_0$ is a constant where $\Phi_0 = 260.15^{\circ}$
is the daytime value of the phase;  $\Phi(t)$ is the measured
phase;  and $\Delta d(t)$ represents the distance along the propagation path covered by the shadow. 

To find out $\widetilde{d(t)}$ we proceed as follows

\begin{itemize}
\item
Compute, with a time cadence of 1 minute, the altitude and azimuth angles of the Sun and the Moon, as
seen by virtual observers located at a set of evenly spaced points
(every 10 km) along the propagation path NDK-RX.
\item
With this, compute the percentage of the solar disk area $A$ covered
by the Moon (via a two circle intersection approach, see 
\ref{apendix:covered} for details),  seen at  each observational
point along the propagation path ($d$), as a function of time ($t$), 
obtaining the curve $A(d,t)$
($A_{jk}$ in  \ref{apendix:matrix}).
Note that, for a fixed time $k$ the function $A(d,t_k)$ represents
 the covered area of the solar disk along the entire path  (this is  equivalent to $A_{k}^{C}$ in  \ref{apendix:matrix}).

\item
For consistency and to facilitate the computations, in this step,
instead $A(d,t_k)$ we  use its percentage (where  100\% is the maximum
  of the solar disk area covered by the moon,  observed  on the entire time
  interval and    over all observational points, later on we
  recover the actual value as weighting process in  \ref{apendix:matrix}).
  Then,
compute the distance $\Delta d(t_k, A_m)$
(or $\Delta d_{km}$ in  \ref{apendix:matrix})
at a set of  evenly spaced percentage values.
As an example Figure \ref{fig:model}b shows
  $A(d,t_k)$ for $t_k =$ 18:00 UT with a continuous line and the
  correspondent $\Delta d(t_k, A_m)$ for $m=60\%$ is marked by the
  horizontal dashed line.

\item
Substitute $\Delta d(t,A_m)$  in Equation \ref{eq:Equation 5} to obtain
$\widetilde{d(t,A_m)}$, this is the  distance illuminated  as
a function of time and percentage (m).
As instance,
Figure \ref{fig:model}c shows two iso-curves at $ m=30 \%$ and
$80 \%$, with long and short dashed lines, respectively; and the colored
contours are the same but for different values of m.

\item
Finally, we define the total change of illuminated distance as the
envelope of all the
iso-curves. This is shown by the continuous line in Figure
\ref{fig:model}c (see  \ref{apendix:matrix} for the
computational details).

\end{itemize}

Once we know the effect of the eclipse on the propagation path, we are able to 
quantify 
the 
increase of  the effective reflection height of the ionosphere caused
by the Moon shadow. In order to find $\Delta z(t)$ (Equation \ref{eq:Equation 3}), we need to do a discrete convolution over two finite sequences:
\begin{equation}
\label{eq:Equation 6}
(f \ast g)[n] = \sum_{m=-M}^{M} f[n-m]g[m],
\end{equation}    
where $f$ and $g$ are one-dimensional input arrays whose convolution
is defined at each overlapping point, 
in this case $f = 1/\widetilde{d(t)}$ and $g =
\widetilde{\Phi(t)}$. The resulting convolution (see Figure
\ref{fig:model}d) is proportional to $C_2 \Delta z(t)$, which is
dimensionless. By normalizing it, we are left with a dimensionless
function of reflection height variation $\widetilde{\Delta z(t)} =
\Delta z(t)/b$, where $b$ is an unknown parameter in units of
km. Putting together equations \ref{eq:Equation 3}, \ref{eq:Equation
  4} and \ref{eq:Equation 5}, we obtain the model of the phase
$\Phi_M$ as:
\begin{equation}
\label{eq:Equation 7}
\Phi_{M}(t) = \Phi_0 - C_1 d(t) \widetilde{\Delta z(t)} b.
\end{equation}    
Still we do not know the value of the parameter $b$. Therefore we
compute $\chi ^2$ between the modelled $\Phi_{M}(t)$ and the measured $\Phi(t)$ for $b \in [0,20]$ km, with a step of 0.1 km. We adopt parameter $b$, where MIN($\chi ^2$) holds. This directly gives us the profile of the reflection height variation at the time of the eclipse,
\begin{equation}
\label{eq:Equation 8}
\Delta z(t) = \widetilde{\Delta z(t)} b.
\end{equation}  
The best fit between 
our model and the measured phase is achieved  when $b = 9.3 \pm 0.1$
km, see Figure \ref{fig:model}e. Meaning that the estimated rise in
reflection height at the time of the maximum eclipse (18:00 UT) on
propagation path NDK-RX, is about $\Delta z_{max} = 9.3 \pm 0.1$ km,
see Figure \ref{fig:model}f. This means that the effective reflection height of the ionosphere was  $z_{max} = 80.0 \pm 0.5$ km at the time of totality. 

As seen in Table \ref{table:dh},
the computed  $\Delta z$ (marked as this work) 
is slightly higher than the similar results obtained with the VLF
technique, although our result is in very good agreement with the
$\Delta z$ computed using rockets.

\begin{table}
  \begin{tabular}{rrcc}
    \hline
     GCP (km) & $\Delta z$ (km) & Date  & Source \\
    \hline
     1245   & 8.00  & 11-Aug-1999 & {\cite{Clilverd:2001}}  \\
     2820   & 6.18  & 30-Jun-1992 & {\cite{MendesDaCosta:1995}}  \\
    3007   &  9.30  & 21-Aug-2017 & This work, Model\\
    3007   &  5-8  & 21-Aug-2017 & This work, Ratio\\
     4800  & 3.00  &  22-Jul-2009 &  {\cite{2019JGRA..124..616V}}  \\
     5700  & 3.00  &  15-Jan-2010 & {\cite{2012P&SS...73..310P}}  \\
     5761  &  3.75  &  1-Aug-2008 &  {\cite{De:2010}}  \\
     $< 10 000$ & 5.14 & 15-Jan-2010 & {\cite{Guha:2012}}  \\
     $>10 000$ & 4.85 & 22-Jul-2009 & {\cite{Guha:2012}}  \\
    Rocket   & 8.00  & 7-Mar-1970 & {\cite{Clilverd:2001}} \\
     Rocket   & 9.00  & 12-Nov-1966  & {\cite{Clilverd:2001}} \\
   \hline
  \end{tabular}
  \caption{Change of $\Delta z$ during solar eclipses measured by
    different authors using VLF waves except for the last two rows, in
    these cases the measurements were made by rockets. This work model
    refers to our eclipse model (Secc. \ref{sec:model}) and Ratio refers to the
  basic comparison between the eclipse effect and the Sunrise and Sunset \ref{sec:obs}.}
  \label{table:dh}
    \end{table}

\section{The Flare Input} \label{sec:flare}
In this section we investigate the effects of the C3.0 flare occurred at the time of
the eclipse,  reaching its maximum X-ray flux at 17:57 UT. This event
presents a unique opportunity to study the ionospheric conditions during the eclipse, i. e., low background radiation, at the excess flux of a solar flare.

We first take the actual data of phase (NDK-RX) and subtract it from
the 6-Gaussian fit, which describes the minimum expected value of the
phase at the time of totality over our propagation path. Thus we
obtain solely the time profile of the  C3.0 flare. 

\begin{figure}
\begin{center}
\includegraphics[scale=0.50]{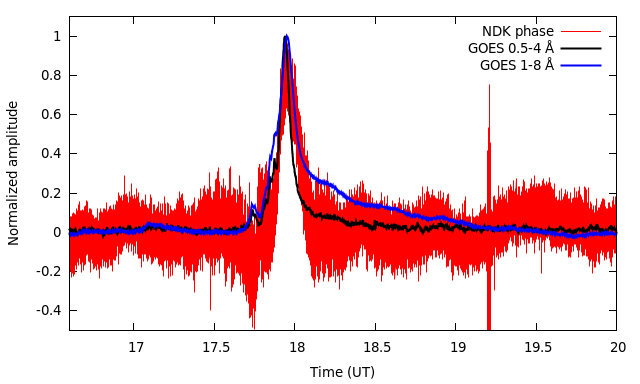}
\caption[f4]{Solar flare X-ray flux from GOES satellite, 0.5-4.0 \AA ~
  (black line) and 1-8 \AA ~ (blue line). Vertical axis represents
 the  normalized amplitude  of GOES flux and the normalized VLF phase (red) for NDK data.}
\label{fig:Figure 5}
\end{center}
\end{figure}

In order to compare the duration and shape of the Solar X-ray flux
excess and the associated ionospheric effect observed through the VLF
data, we have plotted (see Figure \ref{fig:Figure 5}) the normalized VLF phase response
and the Solar X-ray flux from GOES satellite, 1-8 \AA ~ and 0.5-4.0
\AA. We notice a  much more sharper profile than the regular response that usually
appears in the non-eclipse conditions, i. e.,
the phase change (due to a Solar flare) behaves
in strong accordance with  the X-ray flux. This could be attributed
to the different (eclipsed) background conditions in the lower ionosphere, causing an
increase in the effective recombination coefficient.   A detailed
study of this event will be published elsewhere.

We use the ionospheric height profile 
$z = z(t)$, as well as the propagation
distance function $d = d(t)$ (see Section \ref{sec:model})
and  equation \ref{eq:Equation 1}, but in this case, $\Delta \Phi$ is the
phase difference  caused  by the C3.0 flare; $z(t) = z_0 +
\Delta z_e(t)$ corresponds to the ionospheric height profile under the
eclipse conditions; and  $\Delta z_e(t)$ is the eclipsed change of
height (after the convolution of Eq. \ref{eq:Equation 8}, see figure \ref{fig:model}f).

\begin{figure}
\begin{center}
\includegraphics[scale=0.80]{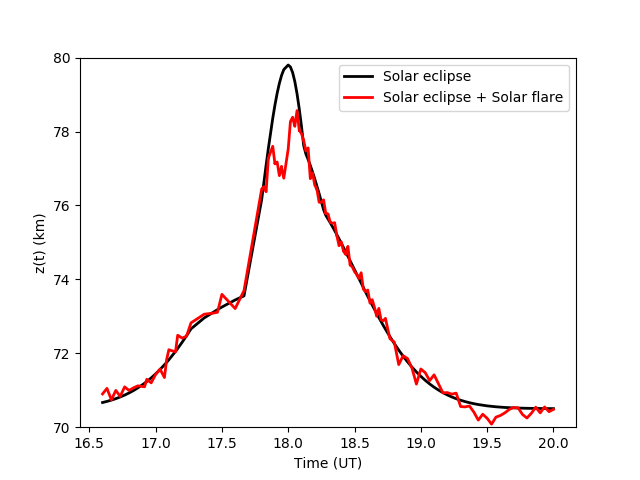}
\caption[f4]{Comparison of ionospheric height of both only the solar
  eclipse and the solar eclipse combined with the solar flare
  input. The solar flare contributes to a maximum change of height by about -3 km $\pm$ 1 km.}
\label{fig:Figure 6}
\end{center}
\end{figure}

We express the flare-induced ionospheric height difference $\Delta
z_f(t)$ to
quantify the 
flare  contribution to the total effective ionospheric height
change and then 
compare the ionospheric height profile during both: the solar eclipse
and the solar eclipse combined with the solar flare input, see figure
\ref{fig:Figure 6}.
Our results are in good agreement with the same obtained by
\citet{2018AdSpR..62..651C}, who measured the signal amplitude of 2
receiving stations in North America, YADA in McBaine and KSTD in
Tulsa.
In the latter they observed similar effects caused by a C3.0 flare.
The authors used the LWPC code and the Wait's two-component D-region
ionospheric model, to numerically reproduce the observed signal
amplitude variation at both receiving locations.
Their altitude profile at maximum electron density for combined
effects (solar eclipse and solar flare) correspond well with our
results, see Figure 8 in \citet{2018AdSpR..62..651C}.  This is
remarkable due to the relative simplicity of our approach.       

\section{Summary and discussion} \label{sec:summary}
We have used the Latin American Very Low Frequency Network at Mexico
(LAVNet-Mex) station to study the effects of the August 21, 2017 total
solar eclipse on the D-region of the Earth's ionosphere.
Our main goal was to estimate the rise of the reflection height at the
time of the eclipse via the measured phase deviation of the received
signal from NDK transmitter in ND, USA (at a fixed frequency of 25.2 kHz).
 We presented the experimental setup used for phase and amplitude measurements of the received VLF waves. 
Also, we  described the so called ``Great American Solar Eclipse'' and
its path across the mainland which crossed almost perpendicularly,
the 3007.15 km long propagation path NDK-RX. As expected, the effect
of the eclipse caused a decrease in both the phase and the amplitude of the
VLF signal.

During the eclipse, a substantial part of the ionosphere above our propagation path
received less ionizing radiation from the Sun. Therefore, to estimate
the rise of the VLF reflection height,
we first approach the problem in a very simple way by 
using as reference the 
 total night/day change (i. e. at the sunrise and the sunset) of 20-30 km, and calculating the ratio between the maximum eclipse phase change and the day-night phase change. 
This accounted for $\sim 25\%$ of the total diurnal average phase change,
equivalent to a rise in reflection height of about 5-8 km. 

Around the totality time (17:57 UT) a C3.0 solar flare  occurred and
was powerful enough to shift the phase and amplitude to more positive
values,  distorting the eclipse pattern. To remove its effects, we
created a model by adding Gaussian functions to better describe the
phase profile during the time of the eclipse totality. 

In 
Section \ref{sec:model} 
we described our model which considers the eclipse geometry to obtain
the area of the solar disk occulted by the Moon as a function of
time. At each time is considered the percentage of the GCP darkening
and obtained the expected values for phase and height variations of
D-region using the \citet{1959ITAP....7..154W}, \citet{1972JATP...34..255D} and \citet{Muraoka:1977} formulations for VLF signal propagation in the Earth-ionosphere waveguide.
In particular, we found that at the time of the eclipse totality (18:00 UT),
the rise in reflection height was  $\Delta z_{max} = 9.3 \pm 0.1$ km. 
Therefore, the effective reflection height of the ionosphere rose to approximately $z_{max} = 80.0 \pm 0.5$ km at the time of totality.

It is important to note that our results of reflection height variation during the eclipse, obtained by two different methods (calculating ratios and eclipse modeling) show a good agreement 
with the results from similar measurements (Table  \ref{table:dh}). 
Particularly, the 
analysis of the ratio between the ionospheric changes during the
eclipse and the day/night transition (Secc. \ref{sec:obs}) is in very good agreement with the results
obtained with similar GCP and analysis method (rows 2 and 4 of
Table \ref{table:dh}). Whereas the result of the eclipse model  (Secc. \ref{sec:model}) is slightly higher
compared with other similar VLF observations, although it is in very good
agreement with the rocket observations (rows 3, 10 and 11 of Table  \ref{table:dh}). 
These differences are expected due to the fact that 
the amount of the reflection height variation during an eclipse depends
on the GCP length, and the relative direction of the GCP and the path
of the eclipse  totality \citep{2019JGRA..124..616V}. In our case, GCP
distance was 3007.15 km and it crossed the path of the total eclipse
almost perpendicularly.

Finally, our model of the C3 X-ray flare observed during the eclipse is
in very good accordance with the results obtained through LWPC simulations reported in \citet{2018AdSpR..62..651C}, for a short-length propagation path. This gives support to the use of this model to study of the eclipsed ionosphere.

\section*{Acknowledgements}
R. Vogrin\v{c}i\v{c} thanks grant Public Scholarship, Development, Disability and Maintenance Fund of the Republic of Slovenia. 
A. Lara thanks UNAM-PASPA for partial support.
J.P. Raulin thanks funding agencies CAPES
(PRINT-88887.310385/2018-00 and Proc. 88881.310386/2018–01) and  CNPq (312066/2016-3) for partial support.
The eclipse timing and location information used in this paper was obtained from the NASA website at \url{eclipse.gsfc.nasa.gov} (Fred Espinak). 
Solar flare soft X-rays fluxes were obtained from the GOES satellite database, website at ftp.swpc.noaa.gov (Prepared by the U.S. Dept. of Commerce, NOAA, 371 Space Weather Prediction Center).
The description of linear convolution used for our eclipse model was
obtained from SciPy website at \url{docs.scipy.org}

\section*{References}
 \bibliographystyle{elsarticle-harv} 
\bibliography{bibtex-records.bib}

\begin{thebibliography}{20}
\expandafter\ifx\csname natexlab\endcsname\relax\def\natexlab#1{#1}\fi
\expandafter\ifx\csname url\endcsname\relax
  \def\url#1{\texttt{#1}}\fi
\expandafter\ifx\csname urlprefix\endcsname\relax\def\urlprefix{URL }\fi

\bibitem[{{Borgazzi} et~al.(2014){Borgazzi}, {Lara}, {Paz}, and
  {Raulin}}]{Borgazzi:2014}
{Borgazzi}, A., {Lara}, A., {Paz}, G., {Raulin}, J.~P., Aug. 2014. {The
  ionosphere and the Latin America VLF Network Mexico (LAVNet-Mex) station}.
  Advances in Space Research 54, 536--545.

\bibitem[{{Chakrabarti} et~al.(2018){Chakrabarti}, {Sasmal}, {Chakraborty},
  {Basak}, and {Tucker}}]{2018AdSpR..62..651C}
{Chakrabarti}, S.~K., {Sasmal}, S., {Chakraborty}, S., {Basak}, T., {Tucker},
  R.~L., Aug. 2018. {Modeling D-region ionospheric response of the Great
  American TSE of August 21, 2017 from VLF signal perturbation}. Advances in
  Space Research 62, 651--661.

\bibitem[{{Clilverd} et~al.(2001){Clilverd}, {Rodger}, {Thomson},
  {Lichtenberger}, {Steinbach}, {Cannon}, and {Angling}}]{Clilverd:2001}
{Clilverd}, M.~A., {Rodger}, C.~J., {Thomson}, N.~R., {Lichtenberger}, J.,
  {Steinbach}, P., {Cannon}, P., {Angling}, M.~J., 2001. {Total solar eclipse
  effects on VLF signals: Observations and modeling}. Radio Science 36,
  773--788.

\bibitem[{Davies(1965)}]{davies65}
Davies, K., 1965. {Ionospheric Radio Propagation}. U.S. Dept. of Commerce,
  National Bureau of Standards, Monograph 80.

\bibitem[{{De} and {Sarkar}(1997)}]{De:1997}
{De}, B.~K., {Sarkar}, S.~K., 1997. {Anomalous behaviour of 22.3 kHz NWC signal
  during total solar eclipse of October 24, 1995.} Kodaikanal Observatory
  Bulletins 13, 205--208.

\bibitem[{{De} et~al.(2010){De}, {De}, {Bandyopadhyay}, {Sarkar}, {Paul},
  {Haldar}, {Barui}, {Datta}, {Paul}, and {Paul}}]{De:2010}
{De}, S.~S., {De}, B.~K., {Bandyopadhyay}, B., {Sarkar}, B.~K., {Paul}, S.,
  {Haldar}, D.~K., {Barui}, S., {Datta}, A., {Paul}, S.~S., {Paul}, N., Oct.
  2010. {The Effects of Solar Eclipse of August 1, 2008 on Earth's Atmospheric
  Parameters}. Pure and Applied Geophysics 167, 1273--1279.

\bibitem[{{Deshpande} and {Mitra}(1972)}]{1972JATP...34..255D}
{Deshpande}, S.~D., {Mitra}, A.~P., Feb 1972. {Ionospheric effects of solar
  flares - IV. Electron density profiles deduced from measurements of SCNA's
  and VLF phase and amplitude.} Journal of Atmospheric and Terrestrial Physics
  34, 255--266.

\bibitem[{{Guha} et~al.(2012){Guha}, {De}, {Choudhury}, and {Roy}}]{Guha:2012}
{Guha}, A., {De}, B.~K., {Choudhury}, A., {Roy}, R., Apr. 2012. {Spectral
  character of VLF sferics propagating inside the Earth-ionosphere waveguide
  during two recent solar eclipses}. Journal of Geophysical Research (Space
  Physics) 117, A04305.

\bibitem[{{Kaufmann} and {Schaal}(1968)}]{Kaufmann:1968}
{Kaufmann}, P., {Schaal}, R.~E., Mar. 1968. {The effect of a total solar
  eclipse on long path VLF transmission}. Journal of Atmospheric and
  Terrestrial Physics 30, 469--471.

\bibitem[{{Mendes Da Costa} et~al.(1995){Mendes Da Costa}, {Paes Leme}, and
  {Rizzo Piazza}}]{MendesDaCosta:1995}
{Mendes Da Costa}, A., {Paes Leme}, N.~M., {Rizzo Piazza}, L., Jan. 1995. Lower
  ionosphere effect observed during the 30 june 1992 total solar eclipse.
  Journal of Atmospheric and Terrestrial Physics 57, 13--17.

\bibitem[{{Muraoka}(1983)}]{1983ZaMM...63..281L}
{Muraoka}, Y., Jan 1983. {A new approach to mode conversion effects observed in
  a mid-latitude VLF transmission}. Zeitschrift Angewandte Mathematik und
  Mechanik 44~(10), 855--862.

\bibitem[{{Muraoka} et~al.(1977){Muraoka}, {Murata}, and {Sato}}]{Muraoka:1977}
{Muraoka}, Y., {Murata}, H., {Sato}, T., Jul. 1977. {The quantitative
  relationship between VLF phase deviations and 1-8 A solar X-ray fluxes during
  solar flares}. Journal of Atmospheric and Terrestrial Physics 39, 787--792.

\bibitem[{Nicolet and Aikin(1960)}]{Nicolet60}
Nicolet, M., Aikin, A.~C., 1960. The formation of the d region of the
  ionosphere. Journal of Geophysical Research (1896-1977) 65~(5), 1469--1483.

\bibitem[{{Pal} et~al.(2012){Pal}, {Maji}, and
  {Chakrabarti}}]{2012P&SS...73..310P}
{Pal}, S., {Maji}, S.~K., {Chakrabarti}, S.~K., Dec 2012. {First ever VLF
  monitoring of the lunar occultation of a solar flare during the 2010 annular
  solar eclipse and its effects on the D-region electron density profile}.
  Planetary and Space Science, 73~(1), 310--317.

\bibitem[{{Raulin} et~al.(2010){Raulin}, {Bertoni}, {Gavil{\'a}n},
  {Guevara-Day}, {Rodriguez}, {Fernandez}, {Correia}, {Kaufmann}, {Pacini},
  {Stekel}, {Lima}, {Schuch}, {Fagundes}, and {Hadano}}]{Raulin:2010}
{Raulin}, J.-P., {Bertoni}, F.~C.~P., {Gavil{\'a}n}, H.~R., {Guevara-Day}, W.,
  {Rodriguez}, R., {Fernandez}, G., {Correia}, E., {Kaufmann}, P., {Pacini},
  A., {Stekel}, T.~R.~C., {Lima}, W.~L.~C., {Schuch}, N.~J., {Fagundes}, P.~R.,
  {Hadano}, R., Jul. 2010. {Solar flare detection sensitivity using the South
  America VLF Network (SAVNET)}. Journal of Geophysical Research (Space
  Physics) 115, A07301.

\bibitem[{{Tereshchenko} et~al.(2015){Tereshchenko}, {Sidorenko},
  {Tereshchenko}, and {Grigoriev}}]{Tereshchenko:2015}
{Tereshchenko}, E.~D., {Sidorenko}, A.~E., {Tereshchenko}, P.~E., {Grigoriev},
  V.~F., Sep. 2015. {Effect of the total solar eclipse of 20 March 2015 on the
  ELF propagation over high-latitude paths}. Geophysical Research Letters 42,
  6899--6905.

\bibitem[{{Thomson}(2010)}]{Thomson:2010}
{Thomson}, N.~R., Sep. 2010. {Daytime tropical D region parameters from short
  path VLF phase and amplitude}. Journal of Geophysical Research (Space
  Physics) 115, A09313.

\bibitem[{{Venkatesham} et~al.(2019){Venkatesham}, {Singh}, {Maurya}, {Dube},
  {Kumar}, and {Phanikumar}}]{2019JGRA..124..616V}
{Venkatesham}, K., {Singh}, R., {Maurya}, A.~K., {Dube}, A., {Kumar}, S.,
  {Phanikumar}, D.~V., Jan 2019. {The 22 July 2009 Total Solar Eclipse:
  Modeling D Region Ionosphere Using Narrowband VLF Observations}. Journal of
  Geophysical Research (Space Physics) 124~(1), 616--627.

\bibitem[{{Wait}(1959)}]{1959ITAP....7..154W}
{Wait}, J., Dec 1959. {Guiding of electromagnetic waves by uniformly rough
  surfaces : Part I}. IEEE Transactions on Antennas and Propagation 7~(5),
  154--162.

\bibitem[{Wait(1968)}]{wait1968mode}
Wait, J.~R., 6 1968. {Mode conversion and refraction effects in the
  Earth‐ionosphere waveguide for VLF radio waves}. Journal of Geophysical
  Research 73~(11), 3537--3548.
\newline\urlprefix\url{https://doi.org/10.1029/JA073i011p03537}

\end{thebibliography}

\appendix
\section{The Covered Solar Disk} \label{apendix:covered}

We present a detailed description of the computation of the covered
solar disk area $A(d,t)$. We use the equations of spherical
trigonometry to calculate the  coordinates of the points
($\Phi_j,\lambda_j$), where \textbf{j = 1, 2, ..., J}, were placed
every 10 km along the propagation path NDK-RX. Each point is at some
distance $d_j$ from the origin, which is set at the receiver station
in Mexico City. Therefore $d_1 = 0$ and $d_J = d$, where $d$ is the
propagation path length. The geographic coordinates of the receiver
and the transmitter at mutual separation $d$ are known. Then: 
\begin{equation}
\label{eqn:Equation 9d}
y = \sin(\lambda_{\mathrm{RX}} - \lambda_{\mathrm{TX}})\cos(\phi_{\mathrm{RX}}),
\end{equation} 
where $\lambda$ is the geographic longitude, $\phi$ is the geographic latitude, $RX$ denotes receiver and $TX$ denotes transmitter, in our case NDK. Similarly it follows
\begin{equation}
\label{eqn:Equation 10d}
x = \cos(\phi_{\mathrm{TX}})\sin(\phi_{\mathrm{RX}})-\sin(\phi_{\mathrm{TX}})\cos(\phi_{\mathrm{RX}})\cos(\lambda_{\mathrm{RX}} - \lambda_{\mathrm{TX}}).
\end{equation} 
We need $y$ and $x$ in order to calculate the Bearing angle $\Theta$,
\begin{equation}
\label{eqn:Equation 11d}
\Theta = \mathrm{atan2}(y,x),
\end{equation} 
where \texttt{atan2} is an inverse trigonometric function of two arguments, which returns a value between $-\pi$ and $\pi$. We calculate the geographic longitude and latitude of the points along the propagation path of length $d$
\begin{equation}
\label{eqn:Equation 12d}
\phi_j = \arcsin\Bigg(\sin(\phi_{\mathrm{TX}})\cos{\Bigg(\frac{d_j}{a}\Bigg)}+\cos(\phi_{\mathrm{TX}})\sin\Bigg(\frac{d_j}{a}\Bigg)\cos(\Theta)\Bigg),
\end{equation} 
where $a$ is the radius of the Earth (6370.0 km). Similarly it follows for the geographic latitude
\begin{equation}
\label{eqn:Equation 13d}
\lambda_j = \lambda_{\mathrm{TX}} + \mathrm{atan2}\Bigg(\sin(\Theta)\sin\Bigg(\frac{d_j}{a}\Bigg)\cos(\phi_{\mathrm{TX}}),\cos\Bigg(\frac{d_j}{a}\Bigg)-\sin(\phi_{\mathrm{TX}})\sin(\phi_{\mathrm{RX}})\Bigg).
\end{equation} 
The length of the propagation path follows as
\begin{equation}
\label{eqn:Equation 14d}
d= \arccos\Bigg(\sin(\phi_{\mathrm{TX}})\sin(\phi_{\mathrm{RX}})+\cos(\phi_{\mathrm{TX}})\cos(\phi_{\mathrm{RX}})\cos(\lambda_{\mathrm{RX}}-\lambda_{\mathrm{TX}})\Bigg)a.
\end{equation}
We are ready to calculate azimuth and altitude of the Sun and the Moon at every point ($\Phi_j$,$\lambda_j$) along the propagation path, for all $t_k$ (every minute of the eclipse, 16:00 UT – 20:00 UT), where \textbf{j = 1, 2, ..., J} and \textbf{k = 1, 2, ..., K}. Using \texttt{astropy.coordinates} library that possess reliable ephemeris \texttt{solar$\_$system$\_$ephemeris.set($'$jpl$'$)}, the coordinates of the Sun and the Moon at an arbitrary time are obtained. Thus we can calculate the angular distance between the centers of the Sun and the Moon as
\begin{equation}
\label{eqn:Equation 15d}
s = \sqrt{(\alpha_M - \alpha_{\odot})^2 + (h_M - h_{\odot})^2},
\end{equation}
where $\alpha$ is azimuth, $h$ is altitude, $M$ denotes the Moon, $\odot$ denotes the Sun. At the time of the eclipse, based from the Earth-ground, radius of the Moon measured $r_M$ = 16.06 arcmin and radius of the Sun measured $r_{\odot}$ = 15.81 arcmin. With known angular separation and radii we calculate the area $A’$ of the solar disk that is covered by the Moon using the circle-circle intersection equation
\label{eqn:Equation 16d}
\begin{multline}
A' = r_M^2
\arccos\left(\frac{s^2+r_M^2-r_{\odot}^2}{2sr_M}\right)+r_{\odot}^2
\arccos\left(\frac{s^2-r_M^2+r_{\odot}^2}{2sr_{\odot}}\right) - \\
\frac{1}{2}\sqrt{(-s+r_M+r_{\odot})(s+r_M-r_{\odot})(s-r_M+r_{\odot})(s+r_M+r_{\odot})},
\end{multline}
where $A'$ has a dimension of (arc min)$^2$. We are interested in the percentage of the covered Sun, thus it follows
\begin{equation}
\label{eq:Equation 17d}
A = \Bigg(\frac{A'}{\pi r_{\odot}^2}\Bigg)100\%.
\end{equation}
We calculate $A$ for every $d_j$ along the propagation path of length $d$ and for every $t_k$ (step of 1 min) in order to obtain a set of curves $A(d,t)$.

\section{Matrix Method} \label{apendix:matrix}
The detailed procedure to model the solar illumination during the
eclipse based in a matrix method is as follows:

\begin{itemize}
\item Let there be \textbf{J} number of points (every 10 km) along the propagation path of a distance d; we denote each point as \textbf{$d_j$} where \textbf{j = 1, 2, ..., J}.
\item Let there be \textbf{K} number of time points from start to end
  (16:00 - 20:00) of the eclipse. 
 Each point is at some distance $d_j$ from the origin, which was set
  at the receiver station in Mexico City. Therefore, $d_1 = 0$ and
  $d_J = d$, where $d$ is the total length of the propagation path (in
  similar way  as in  \ref{apendix:covered}).
\item Let \textbf{$A(d_j,t_k)$} be the calculated area of the covered Sun at an arbitrary point along the propagation path and at an arbitrary time  during the eclipse. We first obtain a matrix \textbf{A} of \textbf{J} $\times$ \textbf{K} elements in this way
\[
A =
\begin{bmatrix}
    A_{11}       & A_{12} & A_{13} & \dots & A_{1K} \\
    A_{21}       & A_{22} & A_{23} & \dots & A_{2K} \\
    \vdots &\vdots &\vdots &\vdots &\vdots \\
    A_{J1}       & A_{J2} & A_{J3} & \dots & A_{JK}
\end{bmatrix},
\]where we simplified $A(d_j,t_k)$ as $A_{jk}$. 
\item We now extract each column in the matrix \textbf{A}. Thus we obtain \textbf{K} number of columns of this form  
\[
A_{k}^{C} =
\begin{bmatrix}
    A_{1k}       \\
    A_{2k}       \\
    \vdots \\
    A_{Jk}     
\end{bmatrix}
\]where \textbf{k = 1, 2, ..., K} and \textbf{C} denotes column. Each $A_{k}^{C}$ can be presented in a graph $A(d,t_k)$ (see Figure \ref{fig:model}b).
\item Let there be \textbf{M} number of possible calculated areas of
  the covered Sun $A$ expressed in $\%$, in our case from 1-99$\%$. We
  calculate the width of each $A(d,t_k)$ curve for each \textbf{m} (see Figure \ref{fig:model}b), where \textbf{m = 1, 2, ..., M}; this gives us $\Delta d(t_k, A_m)$ which can be expressed as a \textbf{K} $\times$ \textbf{M} matrix
\[
\Delta d =
\begin{bmatrix}
    \Delta d_{11}       & \Delta d_{12} & \Delta d_{13} & \dots & \Delta d_{1M} \\
    \Delta d_{21}       & \Delta d_{22} & \Delta d_{23} & \dots & \Delta d_{2M} \\
    \vdots &\vdots &\vdots &\vdots &\vdots \\
    \Delta d_{K1}       & \Delta d_{K2} & \Delta d_{K3} & \dots & \Delta d_{KM}
\end{bmatrix}.
\]
Similarly we simplified $\Delta d(t_k, A_m)$ as $\Delta d_{km}$.
\item We now extract each column in the matrix \textbf{$\Delta d$}. Thus we obtain \textbf{M} number of columns of this form  
\[
\Delta d_{m}^{C} =
\begin{bmatrix}
    \Delta d_{1m}       \\
    \Delta d_{2m}       \\
    \vdots \\
    \Delta d_{Km}     
\end{bmatrix}
\]where \textbf{m = 1, 2, ..., M} and \textbf{C} denotes column. Each $\Delta d_{m}^{C}$ can be presented in a form $\Delta d(t,A_m)$, which we put into Equation \ref{eq:Equation 5} to obtain $\widetilde{d(t,A_m)}$, where \textbf{m = 1, 2, ..., M}.
\item Finally we put weights to each of the \textbf{M} functions $\widetilde{d(t,A_m)}$ in this manner $\widetilde{d(t,A_m)} \rightarrow A_m\widetilde{d(t,A_m)}$, see dashed lines in Figure \ref{fig:model}c. When plotting all weighted $\widetilde{d(t,A_m)}$ together and taking the envelope around the total effective area under these curves, we finally obtain $\widetilde{d(t)}$ (full thick line in Figure \ref{fig:model}c).
\end{itemize}

\end{document}